\documentclass[useAMS,usenatbib]{mn2e}
\usepackage{amsmath}
\usepackage{amsfonts}%
\usepackage{amssymb}
\usepackage{graphicx, natbib, wasysym, subfiles, aas_macros}

\def\r{{\bf r}}
\def\v{{\bf v}}
\def\R{{\bf R}}
\def\F{{\bf F}}
\def\T{{\bf T}}

\def\dst{\displaystyle}

\def\v{{\bf v}}

\def\r{{\bf r}}

\def\B{{\bf B}}
\def\F{{\bf F}}

\newcommand{\parti}[2]{\partial_{#2} #1}
\DeclareMathAlphabet{\textbfsf}{\encodingdefault}{\sfdefault}{bx}{sl}

\usepackage{ulem}

\def\dst{\displaystyle}

\def\parti#1#2{\frac{\partial #1}{\partial #2}}

\def\dst{\displaystyle}
\def\r{{\bf r}}
\usepackage{color}

\title[3D MHD simulations of planet migration]{3D MHD simulations of planet migration in cavities and inner discs of magnetized stars}

\author[Romanova et al.]{\parbox{\textwidth}{M. M. Romanova$^{1,2}$\thanks{E-mail of
corresponding author: \texttt{romanova@astro.cornell.edu}}, A. V.  Koldoba$^{3}$,   G. V.  Ustyugova$^{4}$,  C. C.  Espaillat$^{5,6}$, 
  R. V. E. Lovelace$^{1,2}$}
\vspace{0.4cm}\\
\parbox{\textwidth}{ 
$^{1}$ Department of Astronomy, Cornell University, Ithaca, NY 14853-6801, USA\\
$^{2}$ Carl Sagan Institute, Cornell University, Ithaca, NY 14853-6801, USA\\
$^{3}$ Moscow Institute of Physics and Technology, Dolgoprudny, Moscow Region, Russia \\
$^{4}$ Keldysh Institute for Applied Mathematics, Moscow, Russia  \\
$^{5}$ Institute for Astrophysical Research, Boston University, 725 Commonwealth Avenue, Boston, MA 02215, USA \\
$^{6}$ Department of Astronomy, Boston University, 725 Commonwealth Avenue, Boston, MA 02215, USA \\}}
\date{\today}

\pagerange{\pageref{firstpage}--\pageref{lastpage}} \pubyear{2002}

\begin{document}
\label{firstpage}

\maketitle

\begin{abstract}

\noindent We investigate the Type I migration of planets in low-density cavities and inner discs of strongly magnetized young stars using global three-dimensional (3D) magnetohydrodynamic (MHD) simulations, where the strong magnetic field carves the low-density cavity.  Simulations show that a planet in the cavity migrates inwards up to the radius at which the outer Lindblad resonances are inside the cavity. At smaller radii, the migration stalls. The migration is faster if a star accretes in the unstable regime where the temporary tongues penetrate the magnetosphere.  If a planet is in a highly inclined orbit, it interacts with the disc, and the eccentricity increases due to the Kozai-Lydov mechanism. 

A planet may stop or reverse its migration in the inner disc before entering the cavity.  The magnetosphere interacts with the inner disc,  changing its density distribution such that migration slows down or even be reversed. A tilted magnetosphere also excites density and bending waves in the disc, which may slow down migration and also increase the inclination and eccentricity of the planet.  
When a planet reaches the disc-cavity boundary, it is typically trapped at the boundary  by asymmetric corotation torque. A planet moves together with the boundary when the cavity expands. Overall, a magnetized star provides an environment for slow or reverse migration.

\end{abstract}


\section{Introduction}
\label{sec:Introduction}

Many exoplanets are located in the close proximity of a star, at distances of $\lesssim (0.05-0.1)$ AU (e.g., \citealt{MarcyEtAl2005,Perryman2018}, \textit{exoplanet.eu}). 
It is usually suggested that planets form at large distances from the star and later migrate due to excitation of the Lindblad resonances in the disc (e.g., \citealt{GoldreichTremaine1979}, see also reviews by \citealt{KleyNelson2012,BaruteauEtAl2016}).    
If a star has a strong magnetic field, then it changes the density distribution around, it and the migration pattern of an exoplanet also changes.   

Many young stars have a strong magnetic field, in particular, low-mass Solar-type stars 
(e.g., \citealt{Johns-Krull2007,DonatiLandstreet2009,DonatiEtAl2010}). Strong magnetic field opens a low-density cavity around a star  (e.g.,  \citealt{Konigl1991,Hartmann2000,RomanovaEtAl2003,RomanovaOwocki2015}) where the migration slows down significantly. Therefore, the cavity may be a safe ``haven" for close-in exoplanets (e.g., \citealt{LinEtAl1996,RomanovaLovelace2006}).
However, a few issues should be considered. 

If a planet is in the cavity but is close
to the inner disc, then it continues migrating inward due to Outer Lindblad Resonances located in the disc (e.g., \citealt{LinEtAl1996}). On the other hand, a planet inside the cavity may acquire eccentricity due to the excitation of the Eccentric Lindblad Resonances (ELRs) (e.g., \citealt{GoldreichTremaine1980,ArtymowiczEtAl1991,GoldreichSari2003,OgilvieLubow2003,TeyssandierOgilvie2016}). The growth of eccentricity has been observed in a number of 2D hydrodynamic simulations (e.g., \citealt{PapaloizouEtAl2001,dAngeloEtAl2006,KleyDirksen2006,RiceEtAl2008,BitschEtAl2013,DunhillEtAl2013,
RagusaEtAl2018,DebrasEtAl2021,BaruteauEtAl2021,RomanovaEtAl2023}) and in 3D hydrodynamic simulations (e.g., \citealt{RomanovaEtAl2024}).

On the other hand, a planet on strongly inclined orbit interacts with the disc and the eccentricity and inclination change due to the Kozai-Lidov mechanism (e.g., \citealt{Lidov1962,Kozai1962}) where the remote massive parts of the disc may act as a massive planet or a star \citep{TerquemAjmia2010,TeyssandierEtAl2013}. 3D hydro simulations have shown this mechanism is robust and it operates even when a planet interacts with different parts of the disc \citep{RomanovaEtAl2024}.

A star may accrete in the unstable regime, where matter penetrates through the magnetosphere in temporary equatorial tongues (e.g., \citealt{RomanovaLovelace2006,KulkarniRomanova2008,KulkarniRomanova2009,BlinovaEtAl2016,TakasaoEtAl2022,
ParfreyTchekhovskoy2024,ZhuEtAl2024}). 
This regime is particularly expected at the early stages of stellar evolution when the accretion rate is high and the magnetospheric radius could be smaller than the corotation radius (which is one of the main conditions for the unstable regime, \citealt{BlinovaEtAl2016}). It is important to know the migration rate and the possibility of the planet's survival in the cavity.

The external parts of the magnetosphere may penetrate the inner disc and change the density gradient. The density gradient strongly influences the rate and direction of migration (e.g., \citealt{TanakaEtAl2002,CominsEtAl2016}). If the density gradient is not steep or is positive, then migration stops or reverses. The density gradient is strongly positive at the disc-cavity boundary, and the planet can be trapped at the boundary due to asymmetric corotation torque  (e.g., \citealt{Masset2000,MassetEtAl2006,LiuEtAl2017,RomanovaEtAl2019}). 

A rotating magnetized star with a tilted dipole field excites density and bending waves in the inner disc (e.g., \citealt{TerquemPapaloizou2000}).
A sign of the inner bending wave (a warp) has been observed in spectra of some T Tauri stars (e.g., \citealt{BouvierEtAl1999}). 3D MHD simulations have shown the presence of both types of waves, which have different amplitudes and locations depending on the relative positions of the magnetospheric and corotation radii  
 (e.g., \citealt{RomanovaEtAl2013}). Density waves may change the migration rate and eccentricity of the planet, while bending waves may change the inclination of the planet.

Different authors studied the above processes in the hydrodynamic 2D and 3D simulations, where the low-density
cavity has been modeled in various ways\footnote{For example,  \citet{RagusaEtAl2018,DebrasEtAl2021} supported the density gradient at the disc-cavity boundary by placing a low viscosity in the disc and high viscosity in the cavity.  \citet{RomanovaEtAl2023,RomanovaEtAl2024} developed a long-lasting cavity using stationary initial conditions. They kept a low viscosity in both the disc and the cavity.}.

Compared to prior work, we added for the first time a planet to a global 3D MHD model of the accreting magnetized star, and calculated the planet's orbit inside the cavity carved by the magnetosphere, in the inner disc, and at the disc-cavity boundary.

The plan of the paper is the following. In Sec. \ref{sec:model} we describe our 3D MHD model. In Sec. \ref{sec:cavity} we show results for the planet's migration inside the magnetospheric cavity. 
In Sec. \ref{sec:disc} we show results for the planet migration in the inner parts of the disc and at the disc-cavity boundary. 
We conclude in Sec. \ref{sec:conclusions}. The details of the numerical model are given in Sec. \ref{sec:model-details}

\section{Problem setup and numerical model}
\label{sec:model}

We place a young star of mass  $M_*=M_\odot$ and radius $R_*=2 R_\odot$ in the center of the coordinate system. It
 rotates with the angular velocity $\Omega_*$ (period $P_*$) and has a dipole magnetic field with magnetic moment $\mu_*$ tilted about the rotational axis of the star by a small angle $\theta$. An accretion disc is placed in the equatorial plane of the star and has an inner radius $r_d$ and scale height $h_d$ at this radius. The low-density corona occupies the rest of the simulation region. We derive the initial distribution of the density and pressure from the equilibrium of the gravitational, centrifugal, and pressure gradient forces\footnote{In addition, the corona rotates in cylinders with Keplerian velocities corresponding to velocities in the disc. This condition helps to significantly decrease the initial magnetic braking associated with different angular velocities of the star and the disc.} (see details in \citealt{RomanovaEtAl2002}).
  
During simulations, the disc moves toward the star and is stopped by the stellar magnetosphere at the magnetospheric radius $r_m$ 
where the matter pressure in the inner disc matches the magnetic pressure in the magnetosphere.  The low-density magnetospheric cavity forms at radii $R_*<r<r_{\rm cav}$, where the cavity radius $r_{\rm cav}=r_m$.

A planet is placed either in the cavity, or in the inner disc, or at the disc-cavity boundary. In the cavity, it is placed either in the equatorial plane of the disc,  or at an inclined orbit with inclination angles from $i_0=5^\circ$ up to $i_0=75^\circ$.

We solve a problem in dimensionless form. The reference scale is $r_0=R_*/0.35\approx 2.86 R_*$. 
However, in plots and descriptions, we show 
 distances in radii of the star for convenience.
The size of the simulation region is $R_{\rm out}=10 r_0\approx 28.6 R_*$. 
The reference mass is the mass of the star, $M_0 = M_*=M_\odot$.  
The reference velocity is given by $v_0 = \sqrt{GM_0/r_0}$. The time is measured 
in Keplerian periods of rotation at $r=r_0$: $P_0 = 2
\pi R_0/v_0$. 
 We determine the reference magnetic
field $B_{*0}$ at the equator of the star. Then we vary the magnetic field of the star $B_*=\mu B_{*0}$ by varying the dimensionless parameter $\mu$.  This parameter  
 determines the final size of the magnetosphere in our dimensionless models. 

We take a planet of mass $m_p$ as a part $q_p$ of stellar mass: $m_p=q_p M_*$. We take a planet of Jupiter mass, and
therefore $q_p=10^{-3}$. We also determine the characteristic mass of the disc as a part $q_d$ of the planet's mass: $M_{\rm d0}=q_d m_p$, where the parameter $q_d$ is determined in the code.
 We take $q_d=10$ or $q_d=1$. Then the reference density in the disc:
$\rho_0=M_{\rm d0}/r_0^3=q_d m_p/r_0^3$.

We solve a global system of 3D MHD equations using Godunov-type code on the Cubed Sphere grid in a coordinate system rotating with the star. Cubed sphere represents a cube inflated to a sphere, with coordinates $(x, y, z)$ in each of six blocks of the cubed sphere, where the direction of the $z-$coordinate coincides with the direction of the rotational axis of the star \citep{KoldobaEtAl2002}.  The details of the numerical model are given in Sec. \ref{sec:model-details}.

\begin{table}
\begin{tabular}[]{ ccc }
\hline
Mass of the star                    & $M_*=M_\odot$                             &   $2\times 10^{33}$ g         \\
Radius of the star                 & $ R_*=2\times R_\odot$                  &   $1.4\times 10^{11}$ cm    \\                     
Reference distance                & $r_0=R_*/0.35$                             &   $4\times10^{11}$ cm         \\
Reference velocity                 & $v_0$ [cm s$^{-1}$]                      &   $1.826\times10^{7}$                       \\
Reference period                   & $P_0$ [days]                                  &    1.593                                             \\
Reference density                  & $\rho_0$ [g cm$^{-3}$]                 &  $3.12 \times 10^{-5}q_d (q_p/10^{-3})$   \\
Density at $r=r_{\rm d}$      & $\rho_{\rm d}$ [g cm$^{-3}$]     &  $3.12 \times 10^{-6}q_d (q_p/10^{-3})$   \\
\hline \hline
\end{tabular}
\caption{Reference values. 
 \label{tab:refval}}
\end{table}

\subsection{Calculation of the planet's orbit.}

We calculate the orbit of the planet, taking into account the interaction of a planet with the star and the disc. 
We use the earlier developed approaches  (e.g., \citealt{Kley1998,Masset2000,KleyNelson2012,CominsEtAl2016,RomanovaEtAl2019}).
We find the position $\textbf{r}_{\rm p}$  (the radius vector from the star to
the planet) and velocity $\textbf{v}_{\rm p}$ of the planet at each time step, solving
the equation of motion:
\begin{eqnarray}
 m_{p} \frac{d\textbf{v}_{\rm p}}{dt} =\F_{{\rm star} \to p} + \textbf{F}_{\rm disc\rightarrow p} + \F_{\rm inert}~.
\label{eq:planet}
\end{eqnarray}
The first two terms on the right-hand side represent the gravitational forces acting from the 
star and the disc to the planet, respectively. 
The last term accounts for
the fact that the coordinate system is centered on the star and is not inertial.

A force acting from the star to the planet is:
$$
\F_{{\rm star} \to p} = - \frac{G M_* m_p}{r^3} \r_p~,
$$
and inertial forces:
$$
\F_{\rm inert} = m_p (\Omega_* \times ( \r_p \times \Omega_*) ) + 2 m_p ( \v_p \times \Omega_*)~.
$$
The force from the disc is calculated as follows. We calculate the location of the disc element of mass $dM$ relative to the planet: $\R = \r - \r_p$. Then we calculate the coefficient
$$
C = \left\{
\begin{array}{lcl}
\dst - \frac{G m_p}{R_s^3} \Big( 4 - \frac{3 R^3}{R_s} \Big)~~~~\mbox{at}~~~~R < R_s~,
\\[0.3cm]
\dst - \frac{G m_p}{R^3}~~~~\mbox{at}~~~~R \ge R_s~.
\end{array}
\right.
$$
Here $\dst R_s = 0.8 R_{\rm Hill}$ where $R_{\rm Hill} = r_p ({m_p}/{3 M_*})^{1/3}$ is the Hill radius. We calculate the force by integrating along 
the whole simulation region:
$$
\F_{{\rm disc} \to p} = \int C \R dM~.
$$

The torque acting on the planet from the disc is 
$$
\T = \r_p \times \F_{disc \to p} = \int C ( \r_p \times \R ) dM~.
$$
In the coordinate system rotating with the star, there are no forces of inertia; 
the torque from the star is zero, and the only torque is from the disc matter.

We  calculate the planet's orbital energy and angular momentum
per unit mass using the calculated values of $\textbf{r}_{\rm p}$
and $\textbf{v}_{\rm p}$:
\begin{align}
    E_p = \frac{1}{2}\lvert{\textbf v}_{\rm p}\rvert^{2} - \frac{GM_*}{r_p}
    & & {\rm and}
    & &  {\textbf L_p} = \textbf{r}_{\rm p} \times \textbf{v}_{\rm p}~ .
\end{align}
We use these relationships to calculate the semimajor axis and eccentricity of the planet's orbit at each time step:
\begin{align}
    a_p = -\frac{1}{2}\frac{GM_*}{E_p} & & {\rm and}
    & &
    e_p = \sqrt{1 - \frac{L_p^{2}}{GM_* a_p}} .
\end{align}
The inclination angle of the orbit is
\begin{equation} 
i_p=\arccos{\bigg(\frac{L_{\rm zp}}{L_p}\bigg)} ~,
\end{equation} 
where $L_{\rm zp}-$is the $z-$component of the angular momentum.
All these values and torque on the star are calculated in the coordinate system of the remote observer.

\subsection{Reference models}

In the reference models, we take a planet of mass $m_p= M_J$ ($q_p=10^{-3}$). 
We place a planet in orbits with different semimajor axes from $a_0=2.0$ to $4.0$ for planets inside the cavity, and at larger distances 
from the star for planets inside the disc (see column $a_0$ in Tab.
 \ref{tab:mag-rates}). We take zero initial eccentricity\footnote{In our earlier hydro models
(e.g., \citealt{RomanovaEtAl2024}) we took a small initial eccentricity $e_0=0.02$ which helped to suppress 
the action of the  1:2 eccentric corotation resonance which damps the eccentricity  (e.g., \citealt{GoldreichSari2003,OgilvieLubow2003}).  In current MHD 
simulations, the cavity boundary is not as ordered as in the hydro simulations, and this problem does not exist.}.
 
In models with tilted orbits, we  take initial inclination angles of $i_0=5^\circ, 15^\circ, 30^\circ, 45^\circ, 60^\circ$, and $75^\circ$.   

We take a disc with 
a reference surface density $\rho_d=1$,   viscosity coefficient $\alpha=0.01$,
 and scale height of the disc $h_d=0.01 r_d$ (determined at the inner edge of the disc, $r=r_d$ at $t=0$). 
We vary the magnetic moment of the star. In most models, we take a star with a strong magnetic field (determined by the dimensionless parameter $\mu=2$) which provides a large magnetospheric cavity with a typical value of $r_m\approx (4.5-5.0) R_*$. In models where we investigate migration inside the disc, we take a very small value, $\mu=0.001$. In other models, we take intermediate values: $\mu=0.5$ and $1.5$.  

\begin{figure}
\centering
\includegraphics[width=0.5\textwidth]{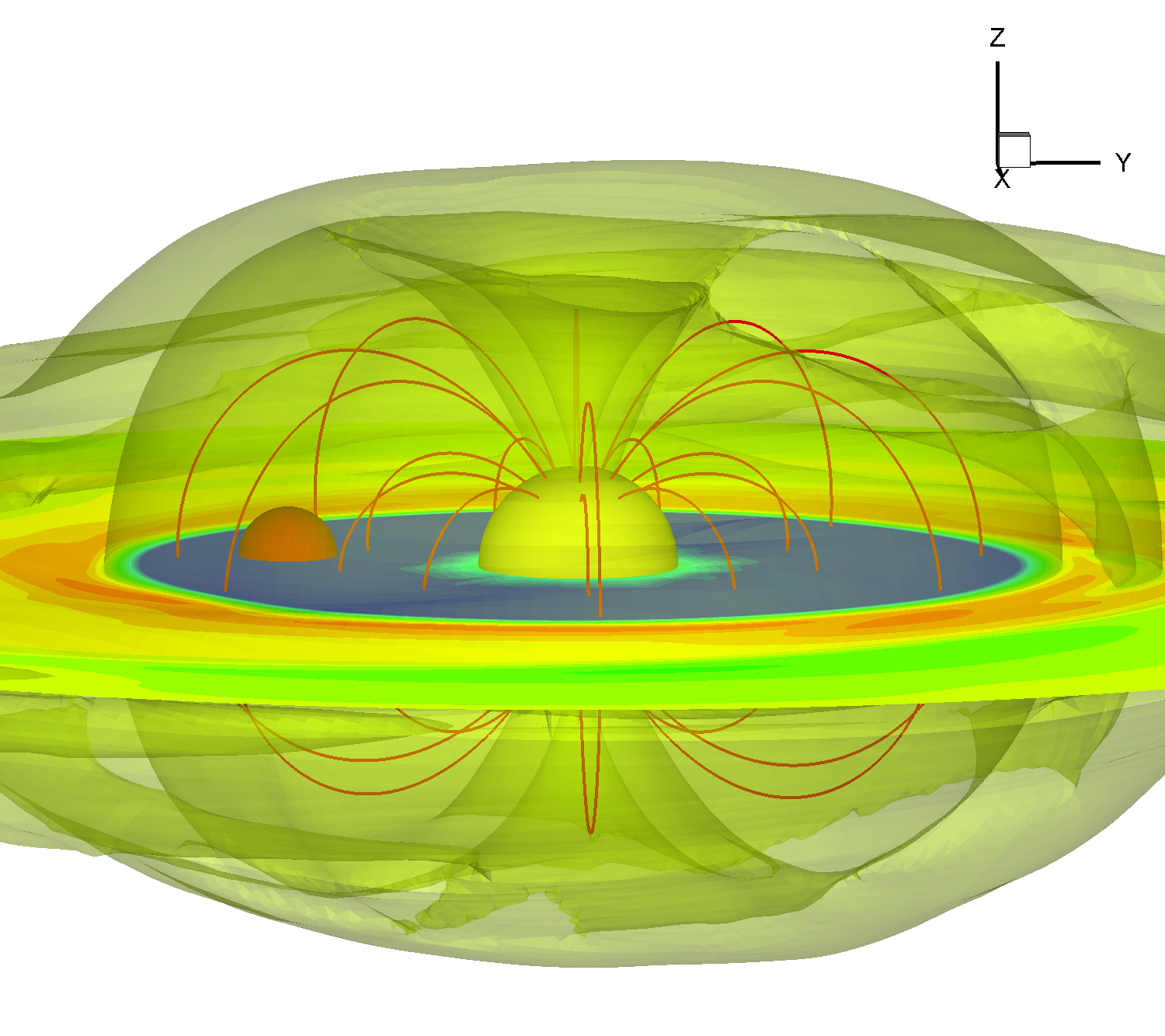}
\caption{An example of a planet (red circle) inside the low-density magnetosphere in the stable regime of accretion in model \textit{Cav2}.
Translucent surfaces show two values of density. Lines are sample magnetic field lines. The equatorial slice shows the density distribution.  
\label{fig:3d-stab}}
\end{figure}

\begin{figure*}
\centering
\includegraphics[height=0.35\textwidth]{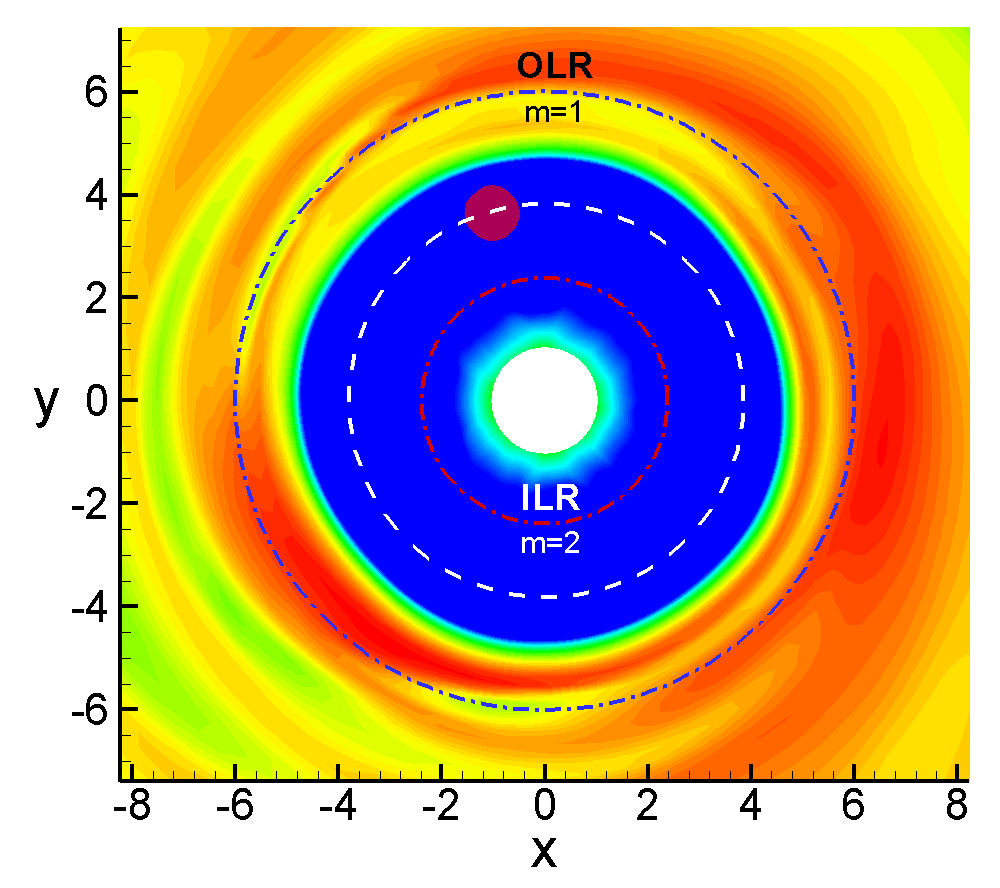}
\includegraphics[height=0.35\textwidth]{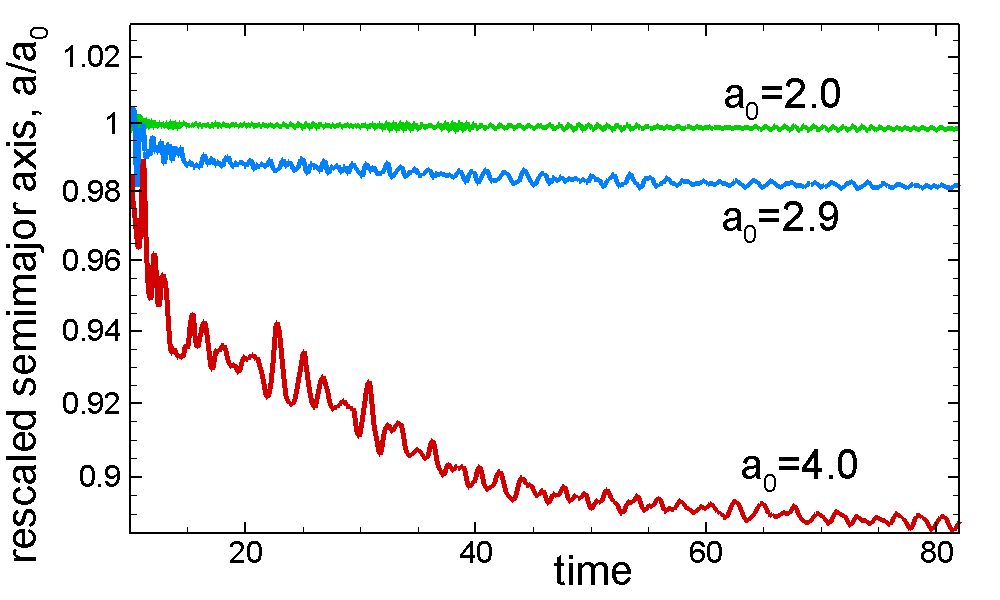}
\caption{\textit{Left panel:} Equatorial slice of density distribution in the model \textit{Cav3}.
 Planet is shown as  red circle. Inner (ILR) and outer (OLR) Lindblad resonances are shown as dashed-dot circles. \textit{Right panel:} Temporal variation of the semimajor axis in models with initial values of $a_0=2.0, 2.86, 4.0$.  In the plot, semimajor axes are scaled such that the initial value equals unity in all models.  
\label{fig:xy-a-stab}}
\end{figure*}

\begin{figure}
\centering
\includegraphics[width=0.49\textwidth]{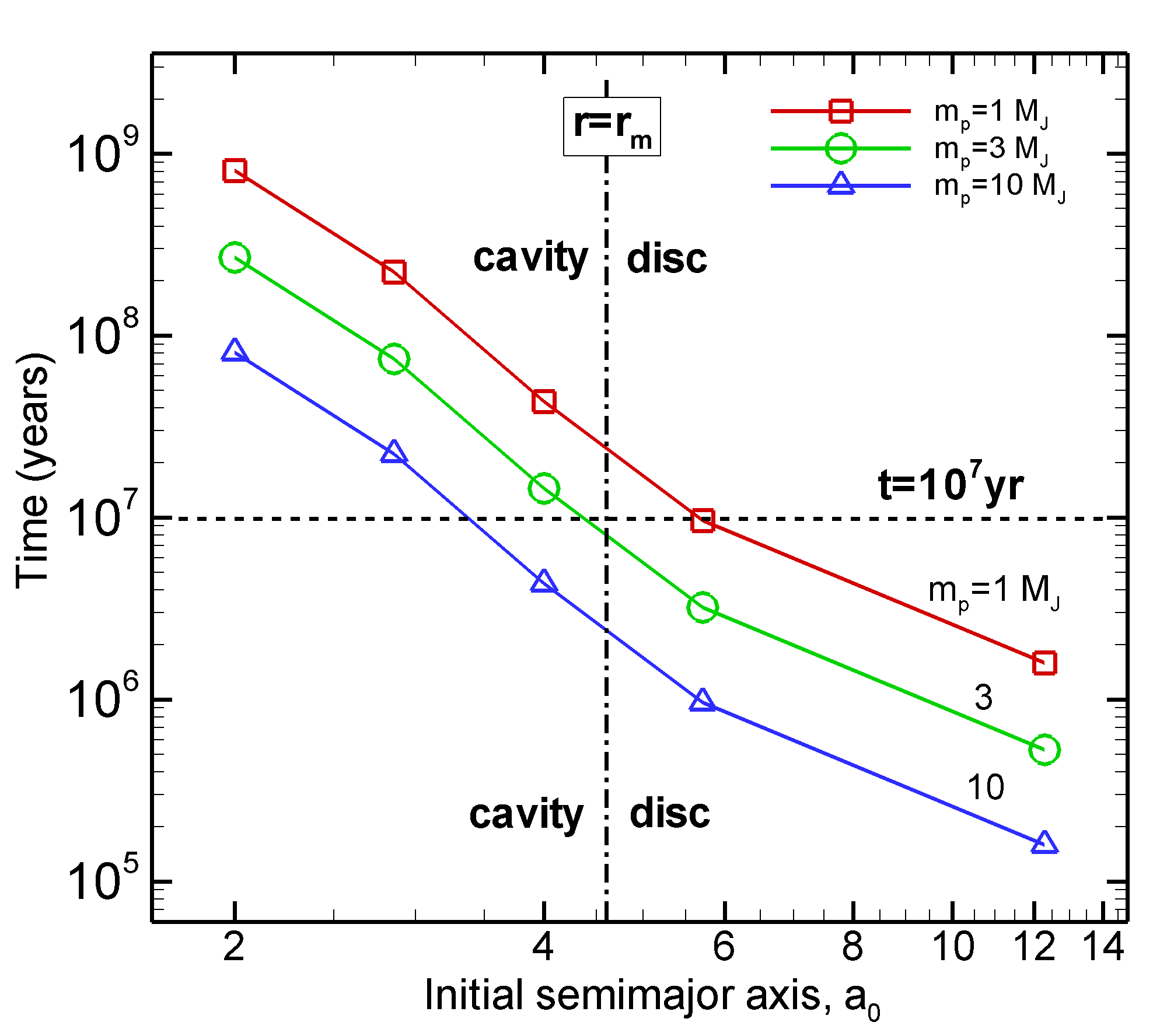}
\caption{Time scales of migration for a planet located in the cavity or the disc as a function of the initial semimajor axis presented in stellar radii. 
The density in the disc and times are rescaled to the disc with the inner density of $\rho=10^{-11}{\rm g cm}^{-3}$. 
Parameters of models are given in Tab. \ref{tab:mag-rates}.
\label{fig:tau-radius}}
\end{figure}

\section{Planet migration in the magnetospheric cavity}
\label{sec:cavity}

\subsection{Basics of planet's migration}
\label{sec:migration}

In case of Type I migration (when a planet does not open a gap in the disc),
the migration is determined by the Lindblad and corotation torques 
\citep{GoldreichTremaine1979,LinPapaloizou1986,Ward1986}. The Lindblad torque is generated when the planet excites $m$-armed waves in the disc with 
frequencies $\omega = m\Omega_{\rm p}$.
The locations of the Lindblad resonances are
\begin{equation}
    r_{\rm LR} = r_{\rm p} \left(\frac{m \pm 1}{m} \right)^{2/3}.
\end{equation}
The outer Lindblad resonances (OLRs, plus sign) exert a negative torque on the planet and ``push'' the planet inward. The inner Lindblad resonances (ILRs, negative sign) exert a positive torque on the planet and ``push'' the planet outward. The total Lindblad torque determines the direction of migration in the absence
of other torques. The OLR, which is most remote from the planet, is at $m=1$ and located at $r_{\rm OLR}=1.59 r_p$.
The most remote ILR is at $m=2$ and located at $r_{\rm ILR}=0.63 r_p$.

There is also a corotation torque associated with co-orbital matter flow (e.g., \citealt{Ward1991,Ward1992}).
The magnitude and sign
of the corotation torque depend on the gradient of the surface density at the corotation radius
 (e.g., \citealt{Masset2001,PaardekooperMellema2006,BaruteauMasset2008,
KleyEtAl2009,MassetBenitez2016}). 
If the density gradient is positive, like at the disc-cavity boundary, then a planet
can be trapped at the boundary due to the asymmetry of the horseshoe orbits and corotation torque, which is directed away from the cavity.

In the disc, the corotation torque dumps the eccentricity. 
 However, in the low-density cavity, 
the corotation torque 
is small, and a planet  
 interacts with the inner disc by Eccentric Lindblad Resonances (ELRs), and the eccentricity increases   (e.g., \citealt{GoldreichTremaine1979,GoldreichTremaine1980,ArtymowiczEtAl1991,GoldreichSari2003,OgilvieLubow2003,
 TeyssandierOgilvie2016}). 
The growth of eccentricity has been observed in 2D and 3D hydro simulations  (e.g., \citealt{PapaloizouEtAl2001,dAngeloEtAl2006,KleyDirksen2006,RiceEtAl2008,DunhillEtAl2013,
 RagusaEtAl2018,DebrasEtAl2021,BaruteauEtAl2021,RomanovaEtAl2023,RomanovaEtAl2024}).

In our models, we consider only migration Type I with no gap opening.
The condition for Type I migration  (e.g., \citealt{LinPapaloizou1993}) is $h_p>R_{\rm Hill}$.
(where $h_p$ is the scale height of the disc at the planet's location), or: 
$h_p/r_p>(q_p/3)^{1/3}$. For $q_p=10^{-3}$ we obtain $h_p/r_p\gtrsim 0.069$. In our model, $h_p/r_p\approx 0.12-0.15$, and the condition for the type I migration is satisfied. 

\subsection{Migration in the cavity: stable regime}

We consider a star with a relatively strong dipole  magnetic field ($\mu=2$) and a magnetic moment tilted about the star's rotational axis at a small angle,  $\theta=5^\circ$. We rotate a star relatively rapidly (corotation radius $R_{\rm cor}=5.7$) so that it accretes in a stable regime through funnel streams, and the magnetospheric cavity has a low density (see models \textit{Cav1, Cav2} in Tab. \ref{tab:mag-rates}). A typical size of the magnetospheric cavity is $r_{\rm cav}=(4.6-5.0) R_*$.

We place a planet into the cavity to orbit with different semimajor axes:  $a_0=2.0, 2.9, 4.0$
 and zero initial eccentricity and inclination: $e_0=0$, $i_0=0$. Fig. \ref{fig:3d-stab} shows a snapshot from simulations of the planet located at the distance $a_0=2.9$. Simulations show that in models \textit{Cav1} and \textit{Cav2}, a planet migrates slowly, while in model \textit{Cav3}, it initially migrates rapidly, then migration slows down (see right panel of Fig. \ref{fig:xy-a-stab}). The reason for the relatively rapid initial migration in model \textit{Cav3} is the fact that a planet interacts with the disc by the OLRs. The left panel of Fig. \ref{fig:xy-a-stab} shows the location of the principal inner and outer Lindblad resonances. At later times, a planet moved to smaller radii, and the action of the Lindblad resonances became weaker. In model \textit{Cav3}, the eccentricity growth due to ELRs was expected but has not been observed. Earlier hydro simulations have shown that several ELRs  were excited at the inner disc, which led to systematic growth of the planet's eccentricity (e.g., \citealt{RomanovaEtAl2023,RomanovaEtAl2024}). We suggest that in the current 3D MHD simulations where the disc-magnetosphere cavity is created by a tilted rotating dipole, the inner disc is not smooth enough for ELRs. Namely, the density inhomogeneities produced by the density and bending waves are much larger than those produced by the ELRs. 
We suggest, however, that the mechanism of eccentricity excitation by ELRs may operate at disc-cavity boundaries developed by other physical mechanisms. 

For comparisons of migration rates, we also calculated two models where a planet is located inside the disc (models \textit{Disc1} and \textit{Disc2}). To remove the effect of the magnetosphere, we used a small magnetic field in the model \textit{Disc1} ($\mu=0.5$) and a very small one in the model \textit{Disc2} ($\mu=10^{-3}$).  

We calculate the migration rate using plots of the semimajor axis ${\rm ln}(a)$ versus time $t$. We take an interval of time in which the curve  is approximately linear,
and calculate the rate of migration $\tau^{-1}$:
\begin{equation}
\tau^{-1} = \frac{d({\rm ln}a)}{dt} \approx \frac{\Delta({\rm ln}{a})}{\Delta t} ~~.
\end{equation}
Then we calculate the characteristic migration time, $\tau$. This value is dimensionless. We convert it to dimensional value $\tau_{\rm dim}=P_0 \tau$ using 
the reference period $P_0=1,593$ days from Tab. \ref{tab:refval}.  Realistic time scales of migration at realistic densities in the disc are millions of years. To significantly decrease the simulation time, we took a disc of high density. After simulations, we rescaled time to more realistic values, using more realistic densities in the disc. Here, we take into account that torque acting on the planet is proportional to the surface density in the disc, and the migration time scale is inversely proportional to the density (e.g., \citealt{GoldreichTremaine1980})\footnote{These dependencies were also tested in 2D and 3D hydro simulations (e.g., \citealt{RomanovaEtAl2023,RomanovaEtAl2024}).}.  For rescaling, we take the dimensional density in our disc $\rho_0$ from Tab. \ref{tab:refval}
\begin{equation}
\rho_{\rm sim}=\widetilde{\rho}\rho_0=3.12\times 10^{-6}{q_d}\bigg(\frac{q_p}{10^{-3}}\bigg)\bigg(\frac{\widetilde{\rho}}{0.1}\bigg)~ {\rm g}{{\rm cm}^{-3}}~,
\end{equation}
where $\rho_0$ is the reference density, and $\widetilde{\rho}\approx 0.1$ is a typical dimensionless density in the disc at the magnetospheric boundary.
 After rescaling to lower densities in the disc, we obtain the expected time scale of the migration:
\begin{equation}
\tau=\tau_{\rm dim}\frac{\rho_{\rm sim}}{\rho_{\rm real}} =3.12\times 10^5\tau_{\rm dim}q_d\bigg(\frac{10^{-11}{\rm g}{{\rm cm}^{-3}} }{\rho_{\rm real}}\bigg)\bigg(\frac{q_p}{10^{-3}}\bigg)~ {\rm yr}~. 
\end{equation}

For rescaling, we take some value of the inner density in the disc, $\rho_d=10^{-11}$ g cm$^{-3}$, which  may correspond to the relatively 
 early stage of the disc-star evolution. 
Tab. \ref{tab:mag-rates} shows migration rates obtained in different models. 

\begin{table*}
\begin{tabular}[]{ccccccccccc }
\hline
Model & $\mu$ &  $\theta$ &  $r_{\rm cor}$    &  $a_0$  & $\Delta t$  & $\widetilde{\tau}$  & $\tau$ (yr)  & $q_d$ & $\tau_{\rm res}$ (yr)   &  location, process  \\
\hline
Cav1 & $2$    &  $5^\circ$ & $5.7$  &  2.0   & $15<t<80$  & $5.95\times 10^4$ &   259.68  & 10 & $8.11\times 10^8$    &  cavity, migration \\       
Cav2 & $2$    & $5^\circ$ & $5.7$   &  2.9   & $15<t<48$  & $5.66\times 10^3$ &  24.70   & 10 &  $7.72\times 10^7$     &  cavity, migration \\ 
  ---     &  $2$ & $5^\circ$ & $5.7$  &  2.9   & $48<t<80$  & $1.64\times 10^4$ &  71.58   & 10 &  $2.24\times 10^8$     &  --- \\ 
Cav3 & $2$  & $5^\circ$ & $5.7$  &  4.0   & $15<t<42$  & $6.12\times 10^2$ &  2.67     & 10 &  $8.34\times 10^6$     &  cavity, migration \\ 
 ---     & $2$  & $5^\circ$ & $5.7$  &  4.0   & $42<t<74$  & $3.18\times 10^3$ & 13.88    & 10 &   $4.34\times 10^7$    &  --- \\ 
Inst1 & $1.5$ & $5^\circ$ & $8.6$ & 4.3   & $15<t<40$  & $1.80\times 10^3$  &  7.86     & 1 &   $2.45\times 10^6$    &  cavity, migration, unstable \\ 
 ---       & $1.5$ & $5^\circ$ & $8.6$ & 4.3   & $40<t<82$  & $3.62\times 10^3$  & 15.80  & 1  &  $4.94\times 10^6$     &  --- 
 \\ 
Disc1&  $0.5$ & $30^\circ$ & $8.6$ & 5.7  & $10<t<200$ & $7.06\times 10^2$ & 3.08   & 10  &  $9.61\times 10^6$     &  disc, migration, waves \\ 
Disc2& $0.001$ & $5^\circ$ & $8.6$ & 14.3  & $10<t<60$  &  $1.16\times 10^2$& 0.51  & 10  &  $1.59\times 10^6$     &  disc, migration \\ 
Boun1& $0.001$ & $5^\circ$ & $8.6$ & 5.7  & $10<t<90$  &  $1.56\times 10^2$& 0.68   & 1  &  $2.12\times 10^5$     &  boundary, trapping \\ 
\hline
\end{tabular}
\caption{Time scales of migration in different models.  The 1st column: the name of the model. The 2nd column: the dimensionless  magnetic moment, $\mu$. The 3rd column: tilt angle of the dipole, $\theta$. The 4th column: the corotation radius, $r_{\rm cor}$. The 5th column: initial semimajor axis of the planet $a_0$.  The 6th column: time intervals $\Delta t$ taken for the calculation of the migration time scales. The 7th column: the  dimensionless migration time scale, $\widetilde{\tau}$.  The 8th column:  the migration time scale in dimensional units (years) for a star with parameters given in Tab. \ref{tab:refval}.
The 9th column: the charactristic disc mass in planet  masses. The 10th column: the rescaled migration time scales in discs with inner density of $\rho=10^{-11}$ g cm$^{-3}$.
The 11th column briefly describes the characteristics of the model.
\label{tab:mag-rates}}
\end{table*}

Fig. \ref{fig:tau-radius} shows migration time scales obtained in models with different semimajor axes $a_p$ and different masses of the planet. We take into account that the torque acting on the planet is proportional to the mass of the planet, and project results for planets with higher mass, 
$m_p=3 M_J$ (circles) and $m_p=10 M_J$ (diamonds).
 Square symbols and top curve show time scales in our main models where $m_p=1 M_J$.  One can see from the figure that for planets with masses $1M_J$ and $3M_J$ located inside the cavity, $r_{\rm cav}\approx 4.6 R_*$, the time scale of migration is longer than the disc evolution time scale $t_{\rm disc}\approx 10^7$ years. These results can be rescaled to larger or smaller values of the density in the disc. If the density is lower, say, $\rho=10^{-12}$ g cm$^{-3}$, then all curves will be a factor of 10 higher and the migration rates lower. We note that the migration rates inside the magnetospheric cavity are expected to be lower than calculated in our models because the density in our cavity has a finite value (approximately 0.01 of the disc density), which provides slow planet migration in the cavity due to interaction with this gas. We expect that realistic migration rates in the cavity are lower than those obtained in our models. On the other hand, the migration rates in the disc may be lower or higher depending on the density and temperature distribution in the disc 
(e.g.,  \citealt{TanakaEtAl2002,BaruteauEtAl2016}).   

\subsection{Migration in the cavity: unstable regime}

\begin{figure}
\centering
\includegraphics[width=0.49\textwidth]{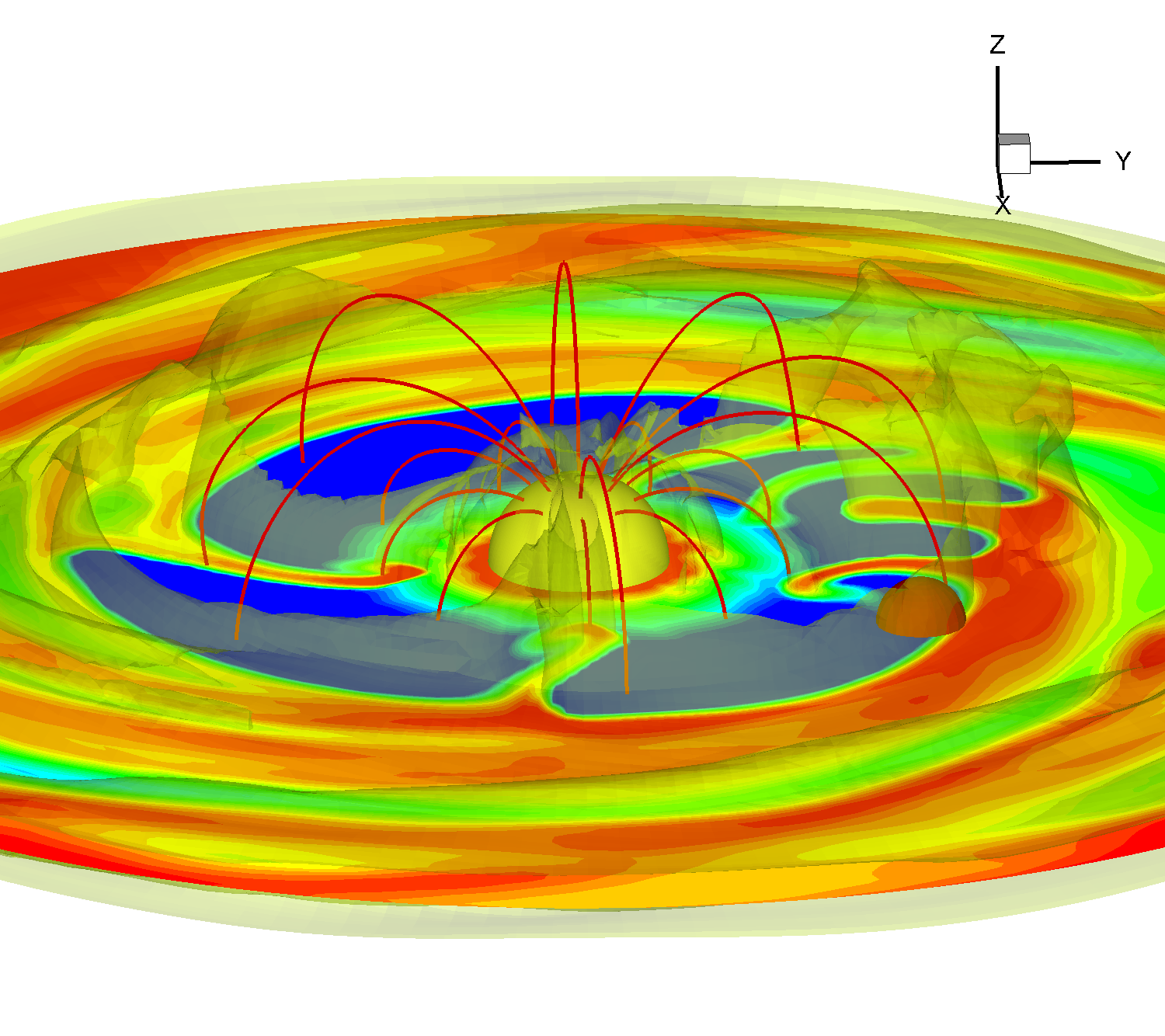}
\caption{Accretion in the unstable regime in the model \textit{Inst1} at time $t=20$. The equatorial slice of the density distribution is shown. Lines are selected field lines. The translucent layer shows one of the density levels.
\label{fig:3d-unstab}}
\end{figure}

\begin{figure*}
\centering
\includegraphics[height=0.35\textwidth]{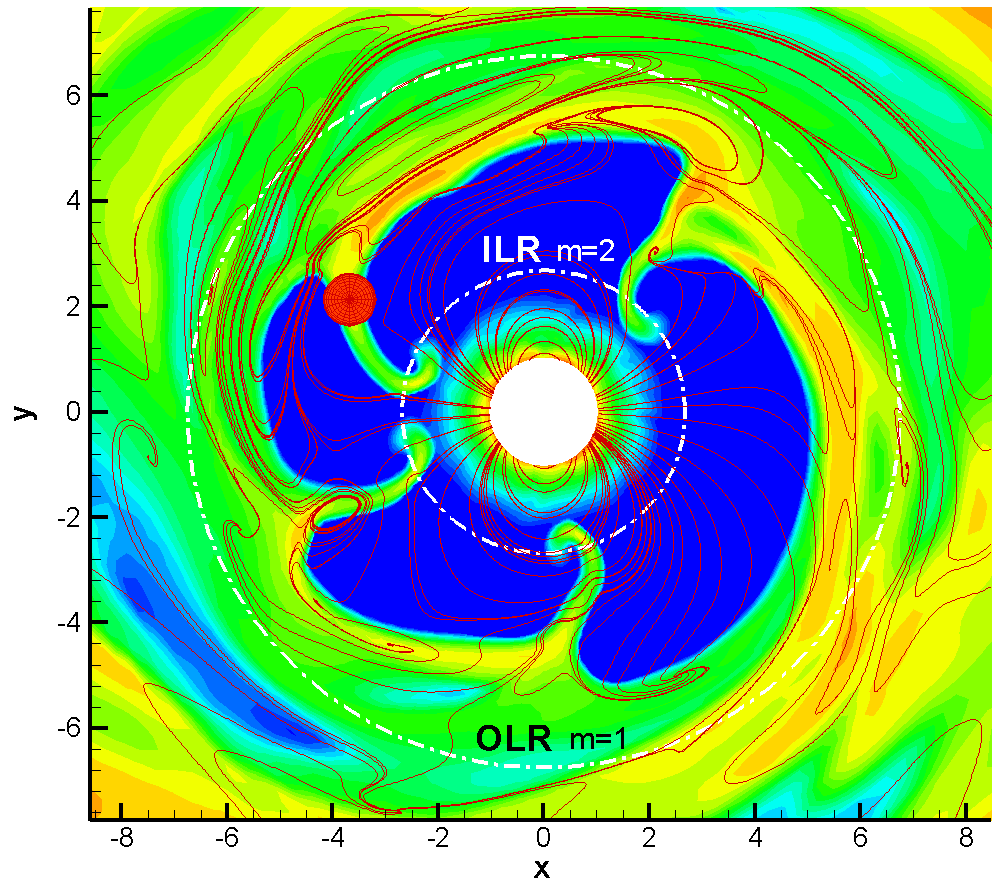}
\includegraphics[height=0.35\textwidth]{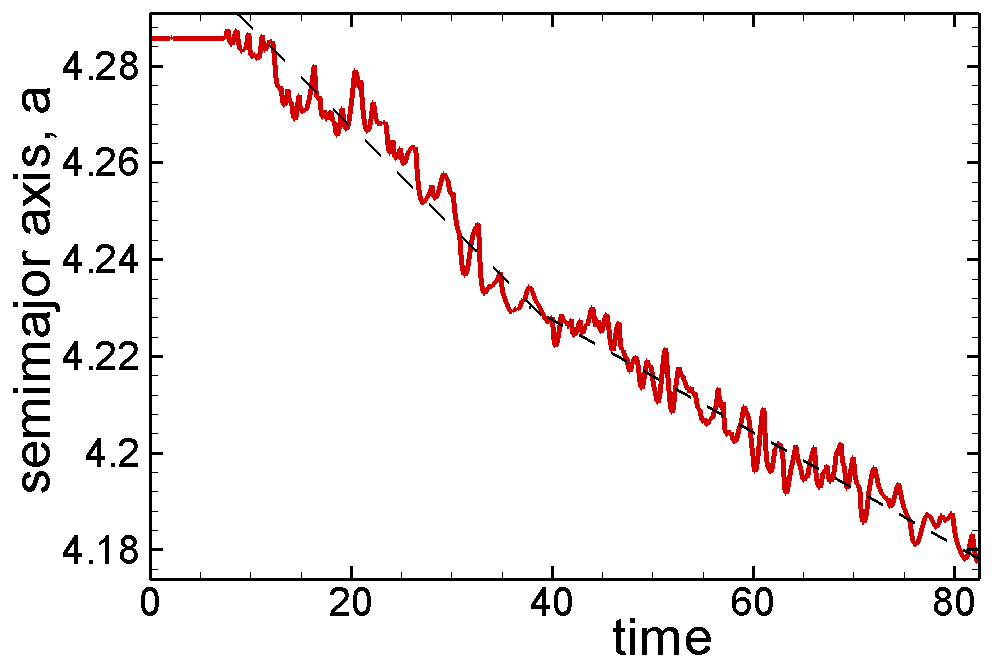}
\caption{\textit{Left Panel:}  Equatorial density distribution at time $t=23.5$ when a planet is passing through unstable tongue (model \textit{Inst1}). Dash-dot lines show the inner and outer Lindblad resonances. The circle shows the position of the planet. \textit{Right Panel:} Temporal evolution of the semimajor axis, a.
\label{fig:unstab-xy-a}}
\end{figure*}

When a star accretes in the unstable regime, the cavity is not empty: a planet interacts with matter in the tongues which penetrate the magnetosphere (see an example in Fig. \ref{fig:3d-unstab}). The left panel of Fig. \ref{fig:unstab-xy-a} shows the equatorial density distribution in a model, where a planet initially located near the disc-magnetosphere boundary and the OLR provides an inward migration of the planet. The right panel of Fig. \ref{fig:unstab-xy-a} shows the variation of the semimajor axis with time. One can see that it fluctuates due to the passage of the planet through tongues. However, overall, a planet migrates inward. The migration rate initially, at $15<t\lesssim 40$, is slightly larger compared with later times,  because the disc initially had a slightly higher density in the inner parts compared with later times.

\begin{figure*}
\centering
\includegraphics[width=0.49\textwidth]{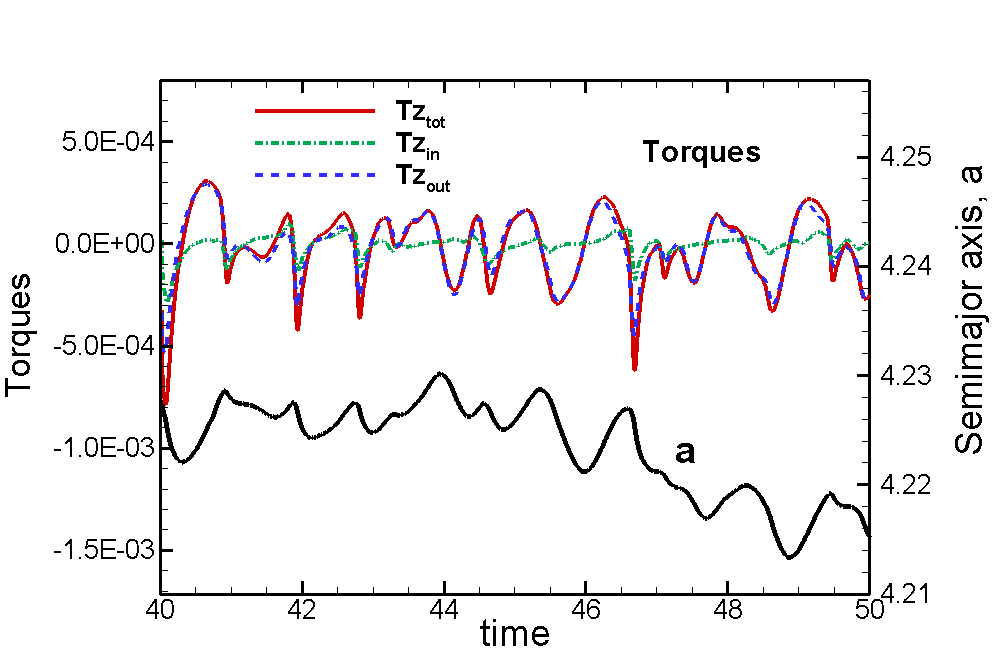}
\includegraphics[width=0.49\textwidth]{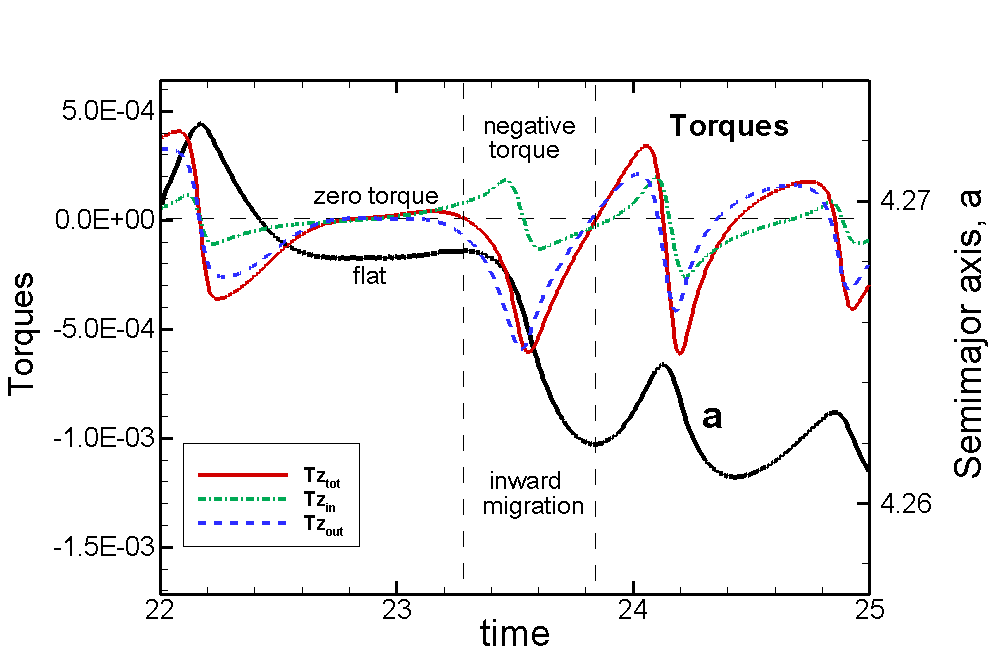}
\caption{\textit{Left panel:} Variation of the semimajor axis (a, black line) and torques acting on the planet in the z-direction   (model \textit{Inst1}). Green and blue dashed lines show inner and outer torques, and the red line shows the total torque. \textit{Right panel:}  The same, but for a smaller interval of time. Times of the negative total torque coincide with times of inward migration. 
\label{fig:unstab-a-torq}}
\end{figure*}

 We analyzed torques acting on the planet from the disc. The left panel of Fig. \ref{fig:unstab-a-torq} shows the evolution of the semimajor axis (a, bottom curve)   in the time interval $40<t<50$ and torques. 
The top curves show torques about the z-axis acting from the disc matter on the planet from matter located inside the orbit (inner torque) $Tz_{\rm in}$ and outside the orbit (outer torque) $Tz_{\rm out}$. These torques are associated with the action of the ILRs and OLRs, which are excited when the planet crosses the unstable tongue. One can see that the total torque, $Tz_{\rm tot}$, is almost entirely determined by the OLRs. The total torque $Tz_{\rm tot}$ (solid line) is associated with the OLR. Both the semimajor axis and torques vary with time due to passage through unstable tongues. 
 The right panel of Fig.  \ref{fig:unstab-a-torq} shows the same but for a smaller interval of time, $22<t<25$. Cross-comparisons of torques and values of the semimajor axis show that when the torque is negative (a planet loses angular momentum), then a planet moves inward and the semimajor axis decreases, and vice versa. Therefore, a planet loses its angular momentum while passing through unstable tongues.

\subsection{Planet on an inclined orbit.
Kozai-Lidov effect}
\label{sec:inclined}

If a planet is on an inclined orbit, then it interacts with the disc, and its inclination and eccentricity may vary due to
Kozai-Lidov effect \citep{Lidov1962,Kozai1962}. In the original version of this effect, it is a three-body problem where a planet interacts with the star and a remote massive planet or another star. \citet{TerquemAjmia2010} have shown that the same effect can be caused by the interaction of the planet with 
the remote parts of the disc, which act as a remote third body. This theoretical prediction has been confirmed by numerical simulations \citep{TerquemAjmia2010,TeyssandierEtAl2013}.  3D hydrodynamic simulations of a planet inside a low-density cavity have shown that the Kozai-Lidov effect also operates when the disc's inner edge is located in a close vicinity of a planet   \citep{RomanovaEtAl2024}. In current 3D MHD simulations, the planet is also in the low-density cavity, and simulations are similar to the 3D hydro simulations \citep{RomanovaEtAl2024}, and the Kozai-Lidov effect was expected.

\begin{figure*}
\centering
\includegraphics[width=1.0\textwidth]{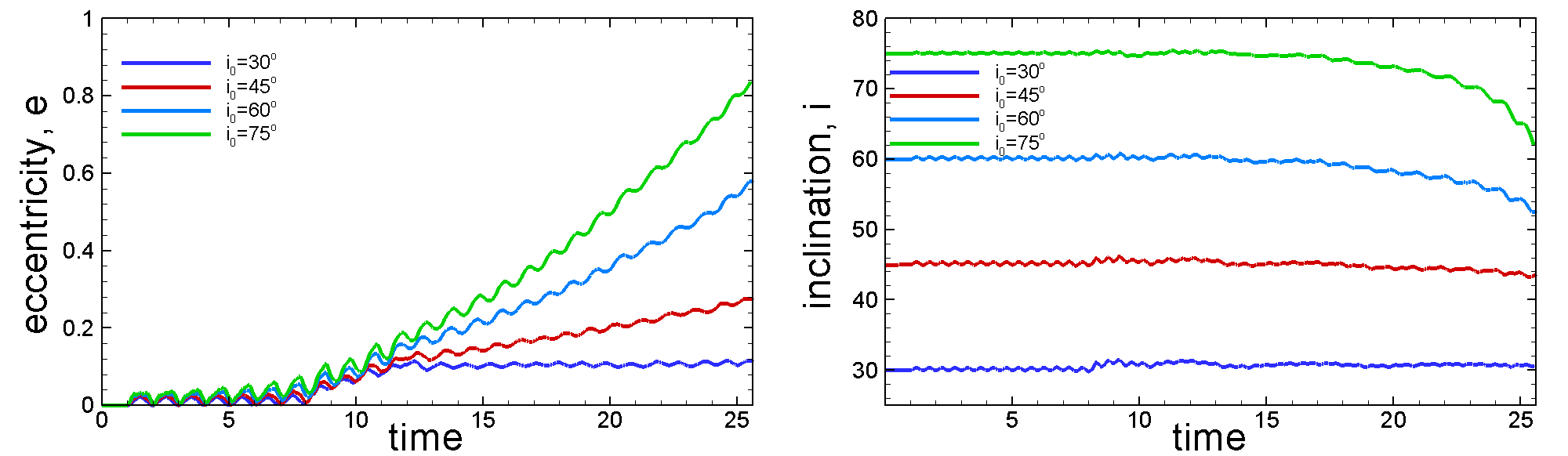}
\caption{Temporal evolution of the eccentricity (left panel) and inclination (right panel) of planets located inside the cavity at 
initial orbits with $a=2.0$ and at different inclination angles $i$. 
\label{fig:kozai-e-i}}
\end{figure*}

To check this effect, we placed a planet on the inclined orbit with different inclination angles $i_0 =30^\circ, 45^\circ, 60^\circ, 75^\circ$.  We took as a base a model \textit{Cav1} where the cavity is large, $r_m\approx 4.6$, and placed a planet  at the orbit with initial semimajor axis  $a_p=2.0$.
Fig. \ref{fig:kozai-e-i} shows that in models with $i_0=45^\circ, 60^\circ, 75^\circ$, the eccentricity increases, while the inclination angle of the orbit decreases. Eccentricity increases more rapidly in models with larger values of $i_0$. 

According to the original theory,
if a planet of mass $m_p$ located at an inclined orbit with semimajor axis $a_p$,
and interacts with a massive object (a planet or a star)  of mass $M_p$ located at the circular orbit of radius $R_p>>a_p$, then  the secular perturbation by the distant companion causes the eccentricity $e_p$ of the inner planet and the mutual inclination $i$ of two orbits  to oscillate in time in antiphase. 
In this situation, the component of the angular momentum 
of the inner orbit perpendicular to the orbital plane, $L_z$ is constant and proportional to
\begin{equation}
L_z\propto \sqrt{1-e_p^2} \cos{i}={\rm const}~.
\end{equation}
This equation shows that the decrease of the inclination angle $i$ leads to the increase of the eccentricity $e_p$, and vice versa.  
As a result, the eccentricity and inclination oscillate in antiphase, and eccentricity can also be pumped to the orbit at the expense of inclination and vice versa.

The maximum value of the eccentricity that can be reached during this process is
\begin{equation}
e_{\rm max} =\bigg(1-\frac{5}{3} \cos^2 i_c\bigg)^{1/2} ~,
\label{eq:emax}
\end{equation}
and therefore the initial inclination $i_0$ should be larger than the critical value $i_c\approx 39^\circ$ which is determined from condition
$
\cos^2{i_c} = 3/5 ~.
$

Our simulations show that at small initial inclination angles, $i_0=5^\circ, 15^\circ, 30^\circ$, the  eccentricity does not increase, which is in accord with the theory.

\begin{figure}
\centering
\includegraphics[width=0.49\textwidth]{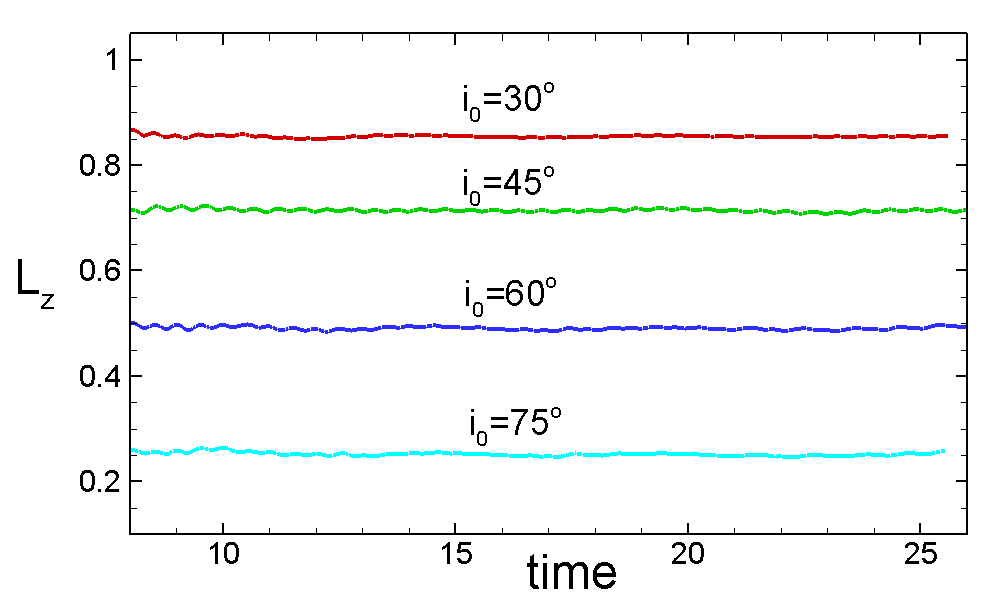}
\caption{Temporal evolution of the z-component of the angular momentum, $L_z$,  for planets located inside the magnetosphere 
with initial orbit with $a=2.0$ and at different inclination angles $i$. 
\label{fig:kozai-Lz}}
\end{figure}

We also calculated the value $\sqrt{1-e_p^2} \cos{i}$ (see Eq. 9 for $L_z$), which is expected to be a constant if Kozai-Lidov mechanism operates. 
Fig. \ref{fig:kozai-Lz} shows that this value is approximately constant in all models. 

A planet in a highly eccentric orbit may hit a star and will not survive (e.g., \citealt{RiceEtAl2008}), unless some other factors help to decrease eccentricity. One of the factors is the tidal interaction with a star, which tends to decrease the eccentricity of the planet's orbit (e.g., \citealt{LinEtAl1996,JacksonEtAl2008}).

Which of these factors will dominate depends on the comparative time scales of the Kozai-Lidov mechanism and other mechanisms, which lead to circularization of the orbit.

\section{Planet in the inner disc}
\label{sec:disc}

A rotating star with a strong magnetic field changes the density distribution in the inner disc, and planets migration may slow down, stop, or reverse. There are a few main factors that may influence the planets' migration: (1) The star's magnetic field partly penetrates the inner disc and changes the density distribution. This may lead to slower migration or its reversal; (2) The magnetosphere with a tilted dipole field excites density and bending waves in the inner disc, which may influence the rate of migration.   
 (3) At the disc-cavity boundary, the density gradient is positive, and the planet may be trapped at the boundary due to asymmetric corotation torque.  

\subsection{Planet trapping at the disc-cavity boundary}

At the disc-cavity boundary, the density gradient is positive and a planet experiences asymmetric corotation torque, which stops inward migration (e.g., \citealt{MassetEtAl2006,LiuEtAl2017}). Global 3D hydrodynamic simulations confirmed planet trapping \citep{RomanovaEtAl2019}.  Here, we investigate this effect in global 3D MHD simulations where the disc-cavity boundary is determined by the magnetosphere of the star.

The disc-magnetosphere boundary differs from the earlier studied disc-cavity boundaries by the fact that it is more dynamic than in prior hydro simulations (where it was almost axisymmetric and quasi-steady). Here, the disc-cavity boundary changes shape with time on a dynamical time scale due to the tilt of the dipole. In addition, the density distribution in the inner disc is not homogeneous due to MRI-driven inhomogeneities and waves excited by the tilted dipole.

\begin{figure*}
\centering
\includegraphics[width=1.0\textwidth]{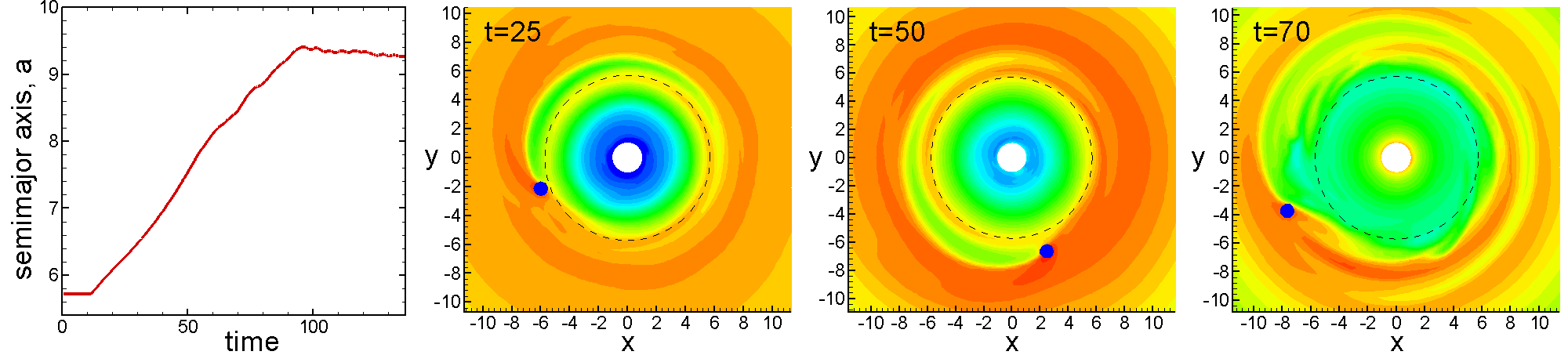}
\caption{A planet is trapped at the disc-cavity boundary due to corotation torque. The left panel shows that the semimajor axis increases with time. The right three panels show that a planet moves outward together with the boundary. 
\label{fig:xy-corot-4}}
\end{figure*}

Here, we demonstrate the action of the corotation torque in the model, where a star initially has a weak dipole magnetic field ($\mu=0.001$) which increases in time due to the azimuthal winding of the magnetic field lines. As a result, the magnetic pressure component in the cavity increases with time, the cavity slowly expands, and a planet moves outward together with the disc-cavity boundary. Fig. \ref{fig:xy-corot-4} shows the equatorial density distribution and position of the planet at different moments. The left panel shows that the semimajor axis systematically increases with time. This numerical experiment shows that a planet can be trapped at the disc-cavity boundary even in cases when the shape of the boundary is not a perfect circle (as in prior 2D and 3D hydro simulations) and has inhomogeneities.

\subsection{Planet migration in the inner disc: waves, MRI inhomogeneities}

A star with a tilted dipole magnetic field generates bending and density waves in the inner disc (e.g., \citealt{RomanovaEtAl2013}).
As an example, we take a model with magnetic moment $\mu=0.5$ tilted at an angle $\theta=30^\circ$ about the rotational axis (model \textit{Disc1}). The rotating tilted dipole produces density and bending waves in the inner disc. At the same time, the magnetic field generates MRI-driven inhomogeneities \citep{BalbusHawley1991,BalbusHawley1998}, which typically represent  azimuthally stretched density enhancements (e.g., \citealt{Hawley2000,Armitage2002,SteinackerPapaloizou2002,RomanovaEtAl2012}). 

\begin{figure*}
\centering
\includegraphics[width=1.0\textwidth]{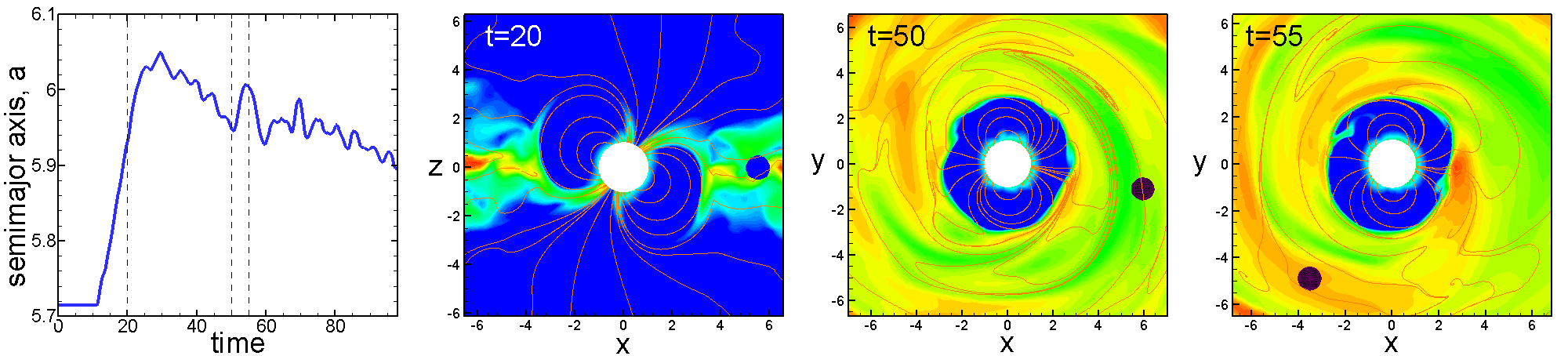}
\caption{An example of planet migration in the inner disc (model \textit{Disc1}). The left panel shows the temporal evolution of the 
semimajor axis.  The 2nd panel shows the poloidal distribution of the density and magnetic field lines at $t=20$. The 3rd and 4th panels show the equatorial density distribution and selected field lines. 
\label{fig:xy-xz-waves-4}}
\end{figure*}

Fig. \ref{fig:xy-xz-waves-4} shows the equatorial density distribution and location of the planet at different moments in time. The right bottom panel shows that  the semimajor axis of the planet's orbit initially increases with time, because initially, both the planet and the inner disc were located at the initial inner radius ($r_d=2$), and the planet experiences the corotation torque due to the high positive density gradient. Subsequently, the matter of the disc moved inward and was stopped by the magnetosphere, while the planet's orbit began to evolve due to inhomogeneities in the disc.  One can see that the semimajor axis of the planet decreases with time. There are fluctuations associated with passage through inhomogeneities in the disc.

The migration time scale is 4-5 times larger compared with the migration time scale in the outer parts of the disc. The
reason is that in the inner disc, the density gradient is not as sharp as in the outer disc due to the penetration of the field lines into the inner parts of the disc and inhomogeneities in the disc.

\subsection{Stochastic growth of inclination and eccentricity}
\label{sec:ecc-incl}

 A planet in the cavity or the inner disc interacts gravitationally with bending waves and gets at each interaction a slight kick
above or below the equatorial plane\footnote{Note that \citet{Terquem2013} studied the growth of inclination  due to the interaction of a planet with remote warped parts of the disc. In their model, the inclination increased due to orbit precession.}. This process is analogous to Brownian motion, where a particle moves in one direction (in spite of random kicks in both directions). Simulations show that inclination gradually increases. In models of planets located inside the cavity, the inclination increases to larger values for a planet placed closer to the inner disc.
The right panel of Fig. 7 shows that the inclination increases up to a 
larger value of $i=3^\circ$ in the model where $a_0=4.0$. 
Inclination increases faster in models where planets are located in the inner disc, where we observed the growth up to $i=7^\circ-10^\circ$.

Eccentricity also increases due to interaction with density waves. This was observed in models of planets in the inner disc. 

Longer simulations are required to understand whether inclination and eccentricity will continue increasing with time.
Both processes may be partly damped by the tidal interaction of the planet with the disc, which tends to decrease them.

\section{Conclusions}
\label{sec:conclusions}

We studied Type I migration of a planet in the environment created by a strongly magnetized rotating young star: in the low-density magnetospheric cavity, in the inner disc, and at the disc-cavity boundary. The main conclusions are the following:

\smallskip

\noindent\textbf{1.} A planet located inside the low-density cavity migrates much more slowly than in the disc. It migrates more rapidly when OLRs are located in the inner disc, but the migration strongly slows 
down when all OLRs are inside the cavity.

\smallskip

\noindent\textbf{2.} If a star accretes in the unstable regime, then a planet experiences temporary negative torque while passing through unstable tongues, and it  migrates more rapidly than in the empty cavity. 

\smallskip

\noindent\textbf{3.} A planet on an inclined orbit interacts with the disc via the Kozai-Lidov mechanism, which causes its eccentricity to increase while its inclination decreases. Eccentricity increases more rapidly and up to higher values in models with a higher initial inclination of the orbit.

\smallskip

\noindent\textbf{4.}  Eccentricity growth due to ELRs has not been observed. We suggest that the disc-cavity boundary created by the tilted magnetosphere and inner disc is too disordered, and the amplitude of resonances created by ELRs is much smaller compared with the amplitude of inhomogeneities in the inner disc.  

\smallskip

\noindent\textbf{5.}  A magnetized star changes the density distribution in the inner disc such that the density gradient becomes less negative or even positive. 
A planet may stop migrating in the inner disc before entering the cavity. 

\smallskip

\noindent\textbf{6.}  At the disc-cavity boundary, the density gradient is strongly positive, and the planet is typically trapped at the boundary due to asymmetric corotation torque. When the cavity expands, a planet moves together with the cavity boundary.  This trapping mechanism is robust  despite of the uneven shape of the disc-magnetosphere boundary.

\smallskip 

\noindent\textbf{7.}  Tilted rotating magnetosphere excites density and bending waves in the inner disc. Interaction with waves slows down the migration, but not significantly.

\smallskip 

\noindent\textbf{8.}  A planet in the inner disc is pushed by bending waves stochastically up or down relative to the equatorial plane, and the inclination of the planet gradually increases. For a planet in the cavity, the inclination increases up to $i=1^\circ-3^\circ$. For a planet in the inner disc, the inclination increases up to $i=7^\circ-10^\circ$. Longer simulations are required to understand whether this mechanism can lead to even larger inclinations of the orbit.   
  
\smallskip
  
If a planet is sufficiently massive, it may open a gap, and the migration is different  (Type II migration). In that case, most of the above processes differ from those considered in this paper.
If the planet migrated to the inner disc, then the cavity carved by the magnetosphere will be combined with the gap opened by the planet, and the cavity will be larger (e.g., \citealt{CridaMorbidelli2007}).
Moreover, a massive planet may disperse a significant part of the inner disc (e.g., \citealt{MonschEtAl2021}). This type of migration near the magnetized star should be studied separately.

\begin{figure*}
\centering
\includegraphics[height=0.28\textwidth]{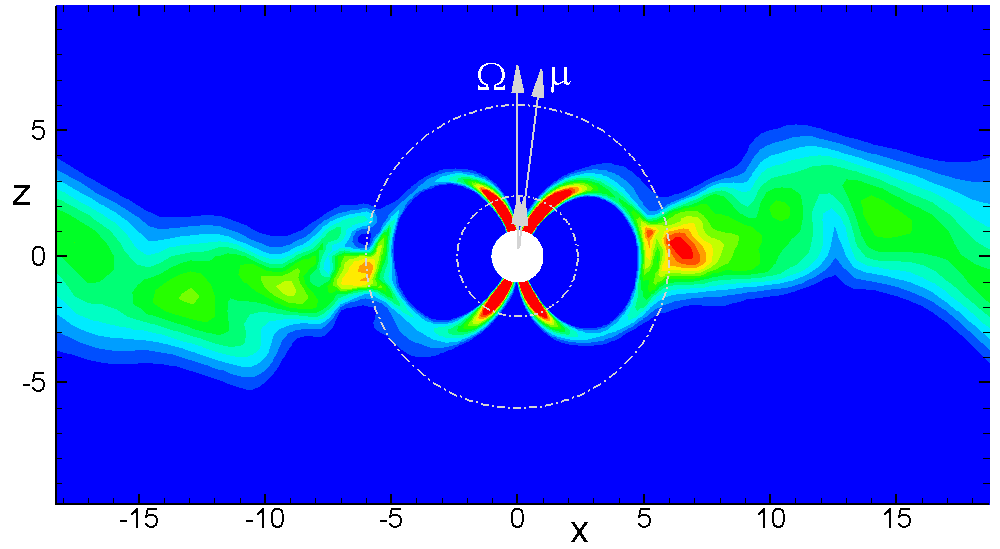}
\includegraphics[height=0.28\textwidth]{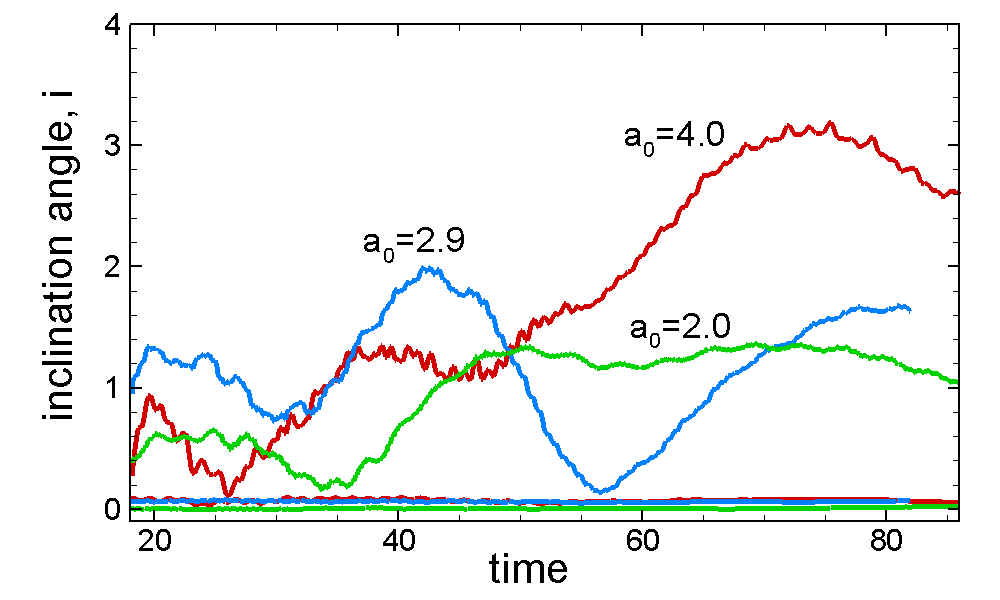}
\caption{\textit{Left panel:} Slice of density distribution in the $\Omega\mu$ ($xz$) plane  in the model \textit{Cav3} at $t=61$.  The inner and outer Lindblad resonances are shown in dash-dot lines. \textit{Right panel:} Temporal variation of the inclination angle and eccentricity 
in models with initial values of the semimajor axis $a_0=2.0, 2.9, 4.0$. 
\label{fig:xz-d2t5r1_4}}
\end{figure*}

\section*{Acknowledgments}
Resources supporting this work were provided by the NASA High-End
Computing (HEC) Program through the NASA Advanced Supercomputing
(NAS) Division at Ames Research Center and the NASA Center for
Computational Sciences (NCCS) at Goddard Space Flight Center. MMR and RVEL were supported
in part by the NSF grant AST-2009820.    CCE was supported by NSF AST-2108446, HST AR-16129, and NASA ADAP
80NSSC20K0451.

\section{Data Availability}

The data underlying this article will be shared on reasonable request to the corresponding author (MMR).

\appendix

\section{Details of the numerical model}
\label{sec:model-details}

We model the evolution of the accretion disc using the 3D MHD  equations of
magnetohydrodynamics, which were solved in the coordinate system rotating with a star with an angular velocity $\Omega_*$.
The entropy balance equation was used instead of the full energy equation
because we do not expect shock waves in this problem. Vector fields were described 
by Cartesian components. The full system of MHD equations can be written in the symbolic form:

\begin{equation}
\parti{\bf U}{t} + \nabla \cdot {\bf F}({\bf U}) = {\bf Q} ~,
\label{eq:hydro}
\end{equation}
where
\begin{equation} 
{\bf U} = \big\{ \rho,~ \rho S,~ \rho \v,~\B \big\}^T~,
\end{equation}
 is the vector of conservative variables. Here $\rho, p-$ are density and gas pressure,
$S\equiv p/\rho^\gamma$ is the entropy function  (we take $\gamma = 5/3$ in all models),
 ${\bf v} = (v_x, v_y, v_z)$ is the
velocity vector, 
 and ${\bf B} = (B_x,
B_y, B_z)$ is the magnetic field.
The vector of fluxes has the form:
\begin{equation}
{\bf F} = \big\{\rho \v,~ \rho S \v,~ {\cal M},~{\cal E} \big\}^T~.
\end{equation}
Here $\cal M$ is the momentum flux tensor, with components $M_{ij} = \rho v_i v_j + \delta_{ij} (p+B^2/8\pi)-B_iB_j/4\pi  - \tau_{ij}$, where
$\delta_{ij}$ is the Kronecker symbol; $\cal E$ is the tensor, with components $E_{ij} = v_j B_i - v_i B_j$;
$\tau_{ij}$ is the tensor of viscous stresses (we take into account only $r\phi-, \theta\phi-$ and
$\phi\phi-$components, recalculated to the Cartesian coordinates). We include a viscosity term, with the viscosity
coefficient in the form of $\alpha-$viscosity, $\nu_{\rm vis}=\alpha c_s H$
\citep{ShakuraSunyaev1973}.
The vector of source terms is
$$
{\bf Q} = \big\{0, 0, -\rho \nabla \Phi + \rho (\Omega_* \times ( \r \times \Omega_*) ) + 2 \rho ( \v \times \Omega_*), ~{\bf 0}
\big\}~,
$$
where $\Phi$ is the gravitational potential of the star.

The MHD equations were integrated numerically using an explicit conservative Godunov-type numerical scheme
\citep{KoldobaEtAl2002}. At both the inner and outer boundaries, most variables $U_j$ are taken to have free boundary
conditions at both the inner and outer boundaries ${\partial U_j}/{\partial r}=0$. At the stellar surface, accreting
gas can cross the surface of the star without creating a disturbance in the flow. These conditions neglect the complex
physics of interaction between the accreting gas and the star. 
The magnetic field is frozen onto the surface of the star. That is, the normal
component of the field, $B_n$, is fixed, while the other components of the magnetic field vary. At the outer boundary,
matter flows freely out of the region. We forbid the back flow of matter from the outer boundary into the simulation
region.

\bibliographystyle{mn2e}

\end{document}